\documentclass[12pt]{article}

\usepackage{amsthm, amsmath, amssymb}
\usepackage{latexsym,amsmath,amsfonts,amsthm,amssymb}
\usepackage{tikz}
\usepackage[dvips]{epsfig}
\usepackage{enumerate}
\usepackage{bbm}
\usepackage{epstopdf}
\usepackage{epsfig}
\usepackage{epstopdf}
\usepackage{epsfig}
\usepackage{import}
\usepackage{wrapfig}
\usepackage{float}
\usepackage{psfrag}
\usepackage{bm}
\usepackage{graphicx}
\usepackage{import}
\usepackage{psfrag}
\usepackage{bm}
\usepackage{graphicx}
\usepackage{graphics}
\usepackage[font={footnotesize},margin=1cm]{caption}

\usepackage[breaklinks=false,pdfpagemode=None,pdfview=FitH,pdfstartview=FitH,citebordercolor={0 0 1},linkbordercolor={0 0 1},urlbordercolor={0 0 1},pagebordercolor={0 0 1},pdfborder={0 0 1}]{hyperref}

\usepackage[normalem]{ulem}

\usetikzlibrary{arrows,shapes}
\usetikzlibrary{decorations.pathmorphing}

\newcommand{\veps}{\varepsilon}

\newcommand{\calJ}{{\mathcal{J}}}

\newcommand{\calO}{{\mathcal{O}}}

\numberwithin{equation}{section}

\title  {
        Many-Body Localization: Stability and Instability
                 }

\author{
Wojciech De Roeck
\\Instituut voor Theoretische Fysica, KU Leuven, Belgium
\\ {\tt wojciech.deroeck@kuleuven.be}
\\and
\\John Z. Imbrie
\\Department of Mathematics,
University of Virginia, USA
\\ {\tt imbrie@virginia.edu}
}
\date{}
\begin{document}
\maketitle
\begin{abstract}
Rare regions with weak disorder (Griffiths regions) have the potential to spoil localization. We describe a non-perturbative construction of local integrals of motion (LIOMs) for a weakly interacting spin chain in one dimension, under a physically reasonable assumption on the statistics of eigenvalues. 
We discuss ideas about the situation in higher dimensions, where one can no longer ensure that interactions involving the Griffiths regions are much smaller than the typical energy-level spacing for such regions. 
We argue that ergodicity is restored in dimension $d > 1$, although equilibration should be extremely slow, similar to the dynamics of glasses. 

 \end{abstract}
\section{Introduction}\label{sec:1}

In recent years, substantial theoretical, experimental, and numerical work has been under way, with a goal of understanding the many-body analog of Anderson localization \cite{Anderson1958,Basko2006,Pal2010}. It is understood that a key feature of many-body localization (MBL) is a failure of thermalization; see \cite{Nandkishore2015} for a review.
In this article we will discuss the status of MBL for strongly disordered spin systems in dimension one and in higher dimensions. 
The situation in one dimension was clarified by work of one of us (JI) \cite{Imbrie2016a,Imbrie2016b} establishing the existence of the MBL phase, under a physically reasonable assumption on level statistics. We will review the key features of this construction to establish a starting point for a discussion of ideas in higher dimensions. In recent work by Huveneers and one of us (WDR) \cite{Roeck2016}, it is argued that MBL breaks down in dimensions $d>1$ due to the destabilizing effects of rare regions. Here we give a simplified version of their arguments, emphasizing the percolation perspective, while making connections to related models that are more analytically tractable. MBL can be understood in one dimension as a failure of resonant regions to form an infinite cluster (``percolate''). Therefore it is natural to investigate the possible breakdown of MBL in higher dimensions by assessing whether resonant regions percolate. 

Many-body localization can be defined in many different ways, but in general it means a violation of ergodicity. Loosely speaking, ergodicity entails the spreading of wavepackets throughout the system.
In a many-body system this would mean throughout the configurations of particles or spins, consistent with a given energy. Thus, in an ergodic system the eigenfunctions take a democratic sampling of all configurations close to a given energy. Spread-out eigenfunctions go hand-in-hand with transport.
At the other extreme, if the eigenfunctions are concentrated about one site (in a single-body system) or about a single configuration (in a many-body system), then this would constitute a failure of ergodicity. 

One way to make the notion of concentration more precise is to define a deformation of the basis vectors used to define the Hamiltonian into the exact eigenvectors. For a single-body system such as the Anderson model on $\mathbb{Z}^d$, the basis vectors are labeled by the sites on the lattice, and with strong disorder (or weak hopping) the Hamiltonian may be diagonalized with a unitary matrix that maps each basis vector to an eigenfunction localized near the associated site. This matrix is typically close to the identity and its matrix elements (the eigenfunctions) decay exponentially with the distance from the site labeling the eigenfunction. We may call such a transformation \emph{quasilocal}. In \cite{Imbrie2016}, such a quasilocal deformation was constructed explicitly.

As we shall see below, one can also define quasilocal deformations for strongly disordered many-body Hamiltonians \cite{Huse2014,Serbyn2013,Bauer2013,Ros2015}. Here, quasilocality means that a rotation involving degrees of freedom in a region of size $R$ should be equal to the identity up to terms exponentially small in $R$. Suppose one has a quasilocal diagonalization of the Hamiltonian.
Then it commutes with the diagonal operators used to define the system
(\textit{e.g.} spin operators). Therefore, if one applies the opposite rotation to these operators, one obtains a complete set of conserved quantities, representing quasilocal deformations of the original operators.  This would constitute an extreme form of non-ergodicity. The existence of these local integrals of motion (LIOMs) is the hallmark of a fully MBL system. See \cite{Imbrie2017} for a review.

\section{Localization in One Dimension}\label{sec:2}

Let us consider a specific model as we delve into the construction of quasilocal unitary operators that diagonalize the Hamiltonian. A key requirement for success of the method is for appropriately defined resonances to form a dilute, nonpercolating set in $\mathbb{Z}$. We consider a random field, random transverse field, random exchange Ising model on
 $\Lambda = [-K, K]\cap \mathbb{Z}$:
\begin{equation}\label{(2.1)}
H = \sum\limits_{i=-K}^{K} h_i S_i^z + \sum\limits_{i=-K}^{K} \gamma_i S_i^x + \sum\limits_{i=-K-1}^{K} J_i S_i^z S^z_{i + 1}.
\end{equation}
This operates on the Hilbert space $\mathcal{H} = \bigotimes_{i\in\Lambda} \mathbb{C}^2$, with 
\begin{equation}\label{(2.2)}
S^z_i =
\begin{pmatrix}
1 & 0\\ 0 & -1
\end{pmatrix}
,
 S^x_i = \begin{pmatrix}
0 & 1\\ 1 & 0
\end{pmatrix}
\end{equation}
operating on the $i$\textsuperscript{th} variable. 
Assume $\gamma_i = \gamma\Gamma_i$ with $\gamma$ small. Random variables $h_i, \Gamma_i, J_i$ are independent and bounded, with bounded probability densities.

The construction proceeds through a sequence of steps wherein rotations are performed on an ever-increasing sequence of length scales.
In the first step, we perform rotations on individual sites.
The only off-diagonal term in (\ref{(2.1)}) is $\gamma_i S^x_i$, which is local. 
We need to identify resonances that may get in the way of using perturbation theory to define the proper rotations on this scale. For now, we only need to look at single-flip resonances.
Let the spin configuration $\sigma^{(i)}$ be equal to $\sigma$ with the spin at $i$ flipped. The associated change in energy is
\begin{equation}\label{(2.3)}
\Delta E_i \equiv E(\sigma) - E (\sigma^{(i)}) = 2 \sigma_i (h_i + J_i \sigma_{i + 1} + J_{i - 1} \sigma_{i - 1}).
\end{equation}
We say that the site $i$ is \textit{resonant} if $|\Delta E_i|<\varepsilon \equiv \gamma^{1/20}$ for at least once choice of $\sigma_{i-1}, \sigma_{i+1}$.
Then for nonresonant sites the ratio $\gamma_i/ \Delta E_i$ is  $ \le \gamma^{19/20}$.
(By using a small power of $\gamma$ for the cutoff on small denominators, we obtain bounds on $n^{\text{th}}$-order diagrams  that are not far off from ``typical'' values $\sim \gamma^n$.)
A site is resonant with probability $\sim 4\varepsilon$. Hence resonant sites form a dilute set  where perturbation theory breaks down.

We use first-order perturbation theory to guide our choice of rotation; the goal at this stage is to diagonalize $H$ up to terms of order $\gamma^2$.
Let $H=H_0 + \calJ$ with $H_0$ diagonal and $\calJ$ off-diagonal. Put
\begin{equation}\label{(2.4)}
\calJ=J^{\textrm{res}}+J^{\textrm{per}}
\end{equation}
where $J^{\textrm{res}}$ contains terms $J(i) \equiv \gamma_i S_i^x$ with $i$ \textit{resonant} (\textit{i.e.}, $\Delta E_i  < \varepsilon$). Then $J^{\textrm{per}}$ contains the remaining ``perturbative'' terms. Put
\begin{equation}\label{(2.5)}
A \equiv \sum_{\text{nonresonant }i} A(i) \text{ with }A(i)_{\sigma \sigma^{(i)}} = \frac{J(i)_{\sigma \sigma^{(i)}}}{E_\sigma-E_{\sigma^{(i)}}}.
\end{equation}
Then we use $e^{-A}$ for a basis change, leading to a new rotated (or renormalized) Hamiltonian:
\begin{equation}\label{(2.6)}
H^{(1)}
 =  e^AHe^{-A} = H+[A,H]+ \frac{[A,[A,H]]}{2!} + \ldots
=  H_0+ J^{\rm res}+J^{(1)}.
\end{equation}

After the change of basis, all the perturbative terms 
$J^{\rm per}$ have been eliminated. The resonant terms
$J^{\rm res}$ are untouched.
The new interaction
$J^{(1)}$ is quadratic and higher order in $\gamma$.
Note that $A(i)$ depends on $\sigma_{i-1}$ and $\sigma_{i+1}$ -- see (\ref{(2.3)}). Thus it may fail to commute with spin operators on neighboring sites. However,
$A(i)$ does commute with $A(j)$ or $J(j)$ if $|i-j|>1$.
Thus we preserve quasi-locality of $J^{(1)}$; it can be written as 
$\sum_g J^{(1)}(g)$, where $g$ is
a sum of connected graphs involving spin flips $J(i)$ and associated energy denominators. 
Specifically, a graph is determined by a sequence of sites $i_0,\ldots,i_n$ such that $\text{dist}(i_p,\{i_0,\ldots,i_{p-1}\}) \le 1$ for $1\le  p \le n$; thus each site in the sequence is equal to or adjacent to one of the previous sites, as required for non-commutativity. We obtain a nonvanishing term in $(\text{ad}\,A)^n J \equiv [A,[A,\ldots,[A,J]\ldots]]$ operating on the spins at those sites.
A graph involving $m$ spin flips has $m-1$ energy denominators and is bounded by $\gamma(\gamma/\veps)^{m-1}$.

Let us define resonant blocks by taking connected components of the set of sites belonging to resonant graphs. We perform exact rotations $O$ in small, isolated resonant blocks to diagonalize the Hamiltonian there. This paves the way for reintegrating such regions into the perturbative framework in subsequent steps.

\begin{figure}[h]
\centering
\includegraphics[width=.9\textwidth]{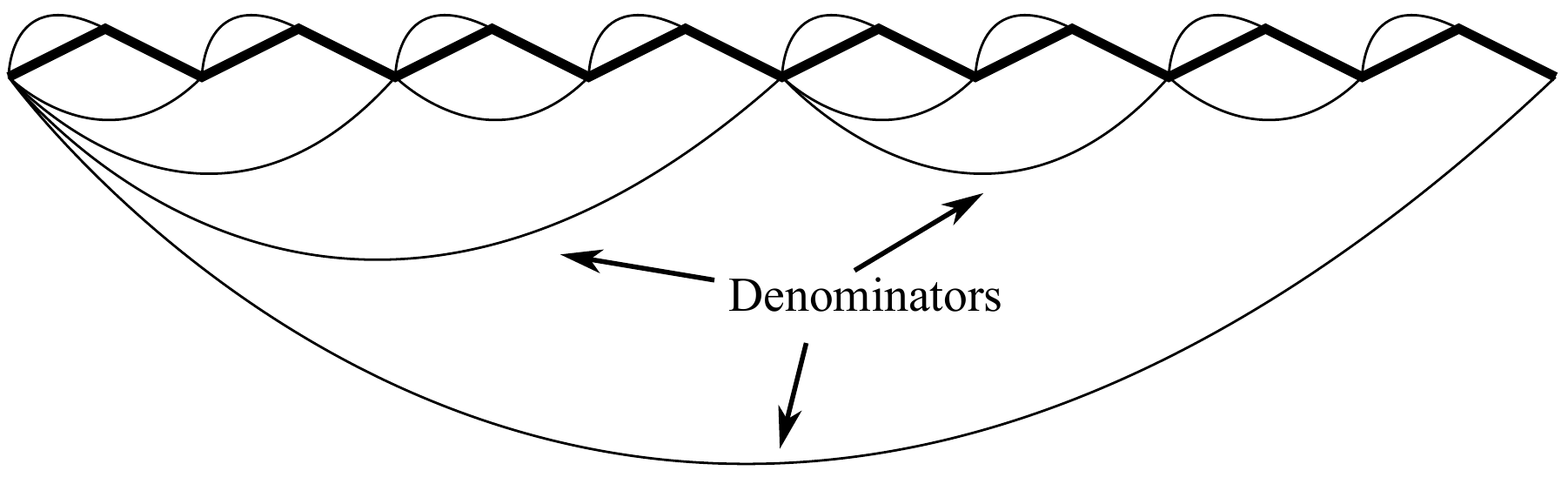}
\caption{Graph contributing to the rotation generator $A^{(k)}_{\sigma \tilde{\sigma}}(g)$.}
\label{fig1}
\end{figure}

Now we may proceed inductively, defining the rotation that is needed to eliminate interactions up to a given order in $\gamma$. We are in effect using Newton's method to solve by successive approximations the problem of diagonalizing $H$. 
Let us define a sequence of length scales $L_k = (15/8)^k$;
then in the $k^{\textrm{th}}$ step we will eliminate interaction terms up to order 
$\gamma^{L_k}$ using operations like (\ref{(2.4)})-(\ref{(2.6)}). 
(A pure Newton's method would lead to remainders $\sim \gamma^{2^k}$; by taking $L_k = (15/8)^k$ we allow for some degeneration of bounds when certain graphs are resummed -- see below.)
The new interaction
$J^{(k)}$ is a sum of connected graphs $J^{(k)}_{\sigma \tilde{\sigma}}(g)$; quasilocality is preserved. In general, a graph of order $L_k$ is defined to be resonant if 
\begin{equation}\label{(2.7)}
A^{(k)}_{\sigma \tilde{\sigma}}(g) \equiv \frac{J^{(k)}_{\sigma \tilde{\sigma}}(g)}{E^{(k)}_\sigma-E^{(k)}_{\tilde{\sigma}}} >
(\gamma/\varepsilon)^{L_k}.
\end{equation}
Then we may perform rotations in the perturbative region (resonance-free region), using 
$\sum_g A^{(k)}_{\sigma \tilde{\sigma}}(g)$ to generate the correct rotation in this step.
The structure of this multiscale perturbation theory is indicated schematically in Fig. \ref{fig1}.

The problem of estimating the probability that $g$ is resonant is less straightforward than in the first step. The idea is to estimate a fractional moment of the graph; a graph of order $L_k$ in $\gamma$ is should have size $\gamma^{L_k}$, and indeed we obtain such a bound on the $s^{\text{th}}$ moment of the graph. In view of the somewhat larger cut-off for resonance in (\ref{(2.7)}), we obtain from a Markov inequality a correspondingly small bound on the probability that $g$ is resonant. Specifically, the probability that $g$ is resonant obeys a bound of order $\veps^{L_k}$.
This makes it possible to sum over $\exp(O(L_k))$ graphs in the associated percolation problem. Complications in this picture arise when graphs visit sites multiple times; this leads to lack of independence in the denominators, which spoils the fractional moment bound. However, by resumming graphs with a substantial fraction of revisits and using previously obtained inductive bounds to cover these cases, one is still able to obtain the requisite exponential bounds on the probability of resonance. Thus we are able to control the ``forward approximation,'' in which it is assumed that each step introduces a fresh random variable.

\subsection{Griffiths regions in $d=1$}

\begin{figure}[h]
\centering
\includegraphics[width=.95\textwidth]{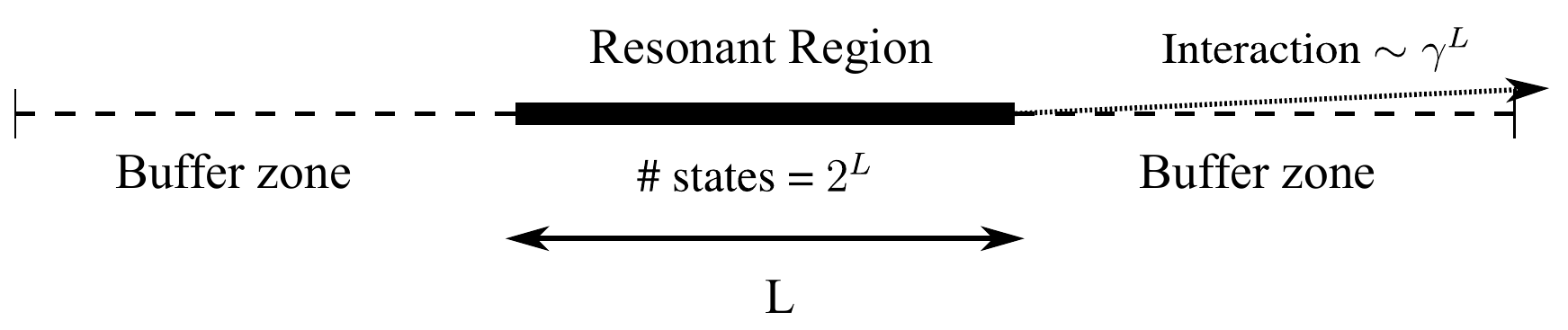}
\caption{Interactions traversing the buffer zone are small enough to compensate for energy denominators of order $2^{-L}$.}
\label{fig2}
\end{figure}

Up to now we have argued why it is likely that the newly generated interaction terms remain non-resonant. 
However, we still need to develop a way to handle those low-probability cases where resonances or near-resonances do occur.
For example, the disorder can be anomalously weak in some region, so that all interactions in this region are resonant. In that case, we need to consider interaction terms connecting a resonant region to its immediate neighborhood or to other resonant regions. Using our assumption on level-spacing statistics, it is possible to show that energy denominators are typically of order $\Delta E \sim 2^{-L}$ for a resonant region of length $L$. However, this remains under control only for interactions spanning a comparable distance. Thus for graphs spanning a distance $L$ we obtain the fundamental requirement for perturbation theory to work:
\begin{equation}\label{(2.8)}
(\text{matrix element})/(\Delta E) \sim \gamma^L/2^{-L} \ll 1.
\end{equation}
As a consequence, a resonant regions of size $L$ needs a buffer zone of size $L$ on either side, and then only terms extending all the way through the buffer zone are considered part of $\calJ^{\text{per}}$ when designing quasilocal rotations as in (\ref{(2.4)})-(\ref{(2.6)}). See Fig. \ref{fig2}.
Lacking control over what happens in the buffer zone, we are forced to do an uncontrolled rotation that diagonalizes the Hamiltonian in the fattened resonant region. However, in one dimension, the buffer zone has volume comparable to that of the resonant region, so the smaller energy denominators $\sim 2^{-3L}$ remain under control.

\begin{figure}[h]
\centering
\includegraphics[width=.9\textwidth]{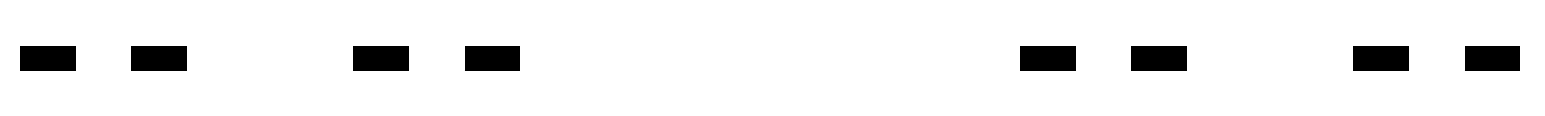}
\caption{Buffer zones on multiple scales lead to extended connectivity for resonant regions.}
\label{fig3}
\end{figure}

When this situation is repeated on multiple scales, one is inevitably led to loosely connected resonant regions, fractal in nature, as indicated in Fig. \ref{fig3}.
Still, the resonances do not percolate. The connectivity function for resonant regions no longer decays exponentially, but it does decay rapidly, faster than any power of the distance. These fractal arrangements of resonant and insulating regions play a role in theories of the MBL transition \cite{Vosk2015,Potter2015,Zhang2016,Agarwal2017}.

\begin{figure}[h]
\centering
\includegraphics[width=.7\textwidth]{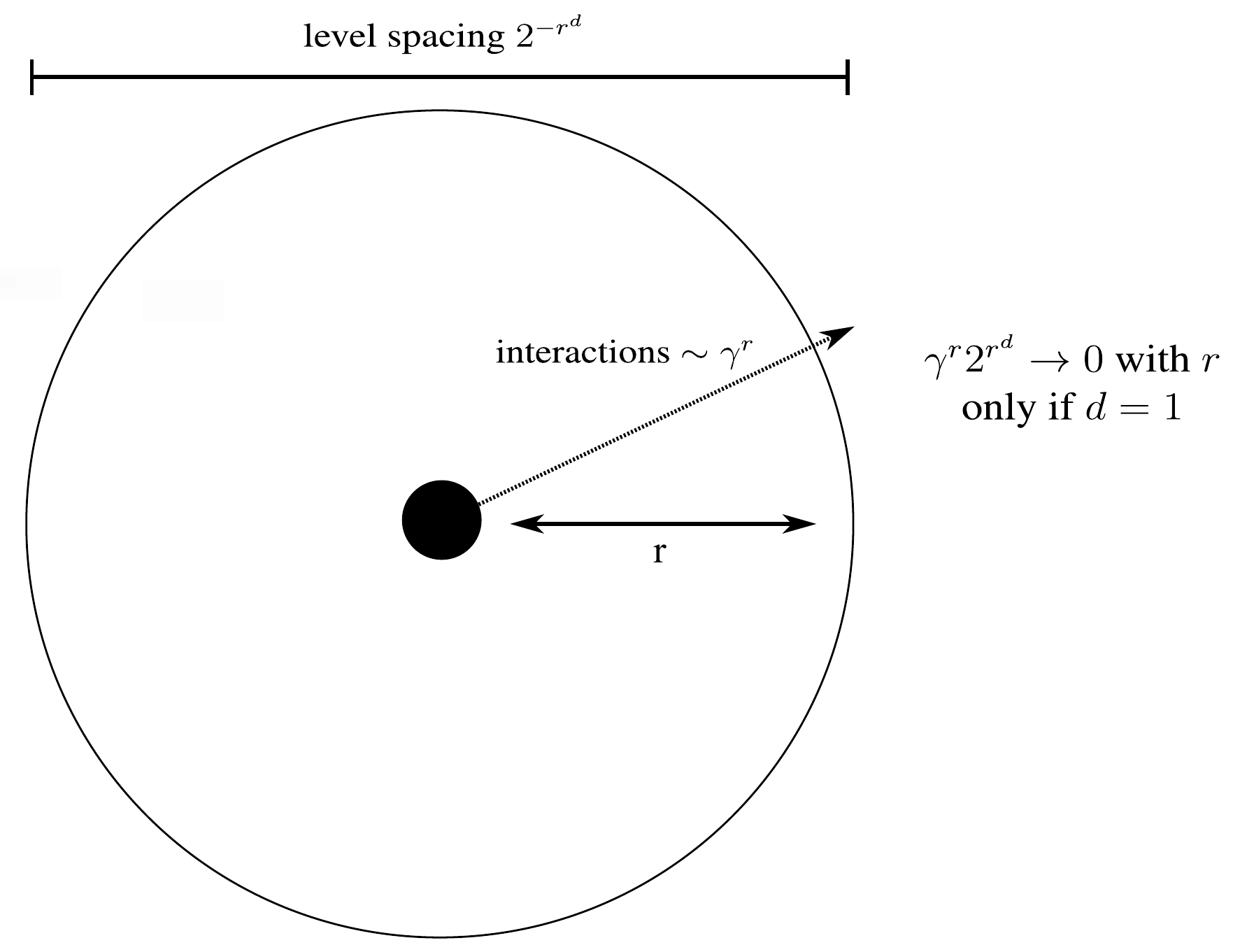}
\caption{In dimension $d>1$ a  buffer zone of width $r$ has volume $\sim r^d$, leading to uncontrollable energy denominators.}
\label{fig4}
\end{figure}

In higher dimensions the buffer zone can have much larger volume, presumably leading to level-spacings that are too small to continue the procedure. See Fig. \ref{fig4}.

\section{Percolation of Resonances in Higher Dimensions}\label{sec:3}

The proof of MBL (summarized above) breaks down in dimension $d > 1$. Is that just a limitation of the method, or does MBL actually break down in higher dimensions? We argue that the resonant regions which proved harmless in $d = 1$ destabilize the MBL phase in higher dimensions. In a random system such as we have been considering here, there are inevitably going to be rare regions of the ``wrong phase,'' otherwise known as Griffiths regions \cite{Griffiths1969}. In such regions, the disorder is anomalously low, presumably leading to thermalization, at least within the low-disorder region or ``bubble''. As we will see, there is a critical bubble size, above which the dimensional scaling of interactions and level-spacings begins to favor resonances. The bubble can then serve as a seed for a cascade of resonant delocalization.

We begin by discussing how percolation of resonances leads to an expectation of ergodicity and delocalization in higher dimensions, referencing a number of related models that shed light on the problem. In subsequent sections, we will review the calculations and numerics of \cite{Roeck2016}, which solidify the basis for the picture presented here.

The first issue we would like to explore is the high degree of connectivity that  a large Griffiths bubble enjoys with its immediate neighborhood. Graphically, it is connected to all if its neighbor sites by interaction terms.
Even though the interactions are small ($\gamma \ll 1$ for the model (\ref{(2.1)}), for a large bubble the number of terms will be much larger than $\gamma^{-1}$, so the perturbative treatment considered above will not lead to small rotation. This is the reason for introducing buffer zones; but now we are trying to understand what happens within the buffer zone. If we consider the bubble as a single vertex in the graph, then it follows that the neighbor sites are only two steps removed from one another. This motivates the consideration of a toy model for resonant delocalization on the complete graph.
Aizenman, Shamis, and Warzel consider a simple model that is relevant for this situation \cite{Aizenman2015}:
\begin{equation}\label{(3.1)}
H_M = -|\varphi_0\rangle \langle \varphi_0| + \kappa_M V
\end{equation}
with
\begin{equation}\label{(3.2)}
\langle \varphi_0 | = (1,1,\ldots,1)/\sqrt{M} \text{ and } \kappa_M = \frac{\lambda}{\sqrt{2 \ln M}}.
\end{equation}
Here $V$ is a random potential, and the kinetic term $-|\varphi_0\rangle \langle \varphi_0|$ connects all of the $M\gg1$ sites to one another.
They find delocalization occurs as predicted by a heuristic condition for resonant delocalization (same as the percolation condition considered above):
\begin{equation}\label{(3.3)}
 (\text{tunneling amplitude})/(\text{level spacing}) \gtrsim 1.
\end{equation}

Let us use this criterion for resonant delocalization to 
examine the status of the graph of resonant transitions in the region surrounding a Griffiths bubble of diameter $L$.
(Bubbles will come in all shapes, but let us consider a roughly spherical bubble with volume of the order of $L^d$.) 
We may consider the most basic transitions that flip a particular spin at a distance $r$ from the bubble, while simultaneously inducing a transition within the Griffiths region. As in the one-dimensional constructions discussed above, we use a basis of eigenstates in the Griffiths region, and in this basis the typical matrix element for such a transition would be of size $\gamma^rN^{-1/2}$, where $N = 2^{L^d}$ is the dimension of the space of states in the bubble. The factor of $N^{-1/2}$ is the appropriate scaling for a local operator $\calO$ in the bubble, since the basis change preserves $\text{tr}\,\calO^*\calO = \sum_{\alpha\beta}|\calO_{\alpha\beta}|^2$. (Assuming the bubble is thermalized, the matrix elements in the new basis should all be roughly of the same size.) Thus the condition for a resonant transition is given by
\begin{equation}\label{(3.5)}
 (\text{tunneling amplitude})/(\text{level spacing})  = \frac{\gamma^{r}N^{-1/2}}{N^{-1}} = \gamma^{r}2^{L^d/2} \gtrsim 1.
\end{equation}
We see that the resonance condition (\ref{(3.5)}) should be satisfied for virtually every spin in a buffer zone defined as $r \le r(L) \approx \tfrac{1}{2}L^d/|\log \gamma|$. That is, the energy of flipping the spin can be precisely matched to the energy of some transition within the bubble. 
The degree of precision required for resonance varies with $r$ in proportion to the matrix element, and thus when $r$ increases past 
$r(L)$, it becomes smaller than the typical level spacing. Nevertheless, we see that virtually every spin configuration within the buffer zone of width $r(L)$ can be reached from any other spin configuration by a sequence of resonant transitions (consistent with conservation of energy, within the energy window of size $\sim \gamma^r$).

Note that for large enough $L$ we have $r(L) \gg L$ for $d>1$. In one dimension, $r(L)$ is of the same order as $L$, as we have discussed already. The dimensional argument drives the whole analysis; it may be compared with the Imry-Ma argument comparing $L^{d/2}$  fluctuations of the bulk free energy with $L^{d-1}$ surface energies for the random-field Ising model \cite{Imry1975}. Although the validity of the Imry-Ma argument was questioned \cite{Parisi1979}, a rigorous analysis \cite{Imbrie1984,Imbrie1985} demonstrated the validity of the scaling picture, employing the Imry-Ma argument in an induction on length scales.
In assessing the viability of the MBL state in higher dimensions, we do not claim the scaling argument is definitive on its own, but when employed in a length-scale induction and buttressed by further analysis, it deserves to be considered the ``default'' prediction, barring convincing evidence to the contrary.

In general, when conditions for resonant tunneling are satisfied, one should expect that all the configurations reachable by resonant transitions should be represented approximately equally in the eigenfunctions. Thus this picture predicts a sort of ``thin ergodicity'' out to a distance at least $r(L)$, meaning that all configurations within an energy window (whose width decreases exponentially with $r$) should be roughly equally represented in the eigenfunctions.

In fact, larger neighborhoods of the Griffiths region should be affected similarly, with ``thin ergodicity'' extending to arbitrarily large radii, in the idealized case of a single Griffiths bubble in the whole of $\mathbb{Z}^d$. To see this, we need a bootstrap (or inductive) argument. In order to match up the transition energy to the requisite accuracy for a spin at a large distance $r$ from the bubble, we need sufficient ``digits of precision'' to represent the energy difference. Thus, in order to find a transition that resonates with the distant spin, we need to enlist a comparable number $O(r|\log \gamma|)$ spins in the vicinity of the bubble. We may choose to organize the induction by considering whether thin ergodicity out to a distance $r/2$ from the bubble will thermalize spins out to a distance $r$. Repeating the calculation in (\ref{(3.5)}) with $N \rightarrow N(r/2) \equiv 2^{(r/2)^d}\gamma^{r/2}$ (to account for the thinness of the percolation cluster), we obtain the condition
\begin{equation}\label{(3.6)}
 (\text{tunneling amplitude})/(\text{level spacing})  = \frac{\gamma^{r}N(r/2)^{-1/2}}{N(r/2)^{-1}} = \gamma^{r/2}2^{(r/2)^d/2} \gtrsim 1.
\end{equation}
Thus we see that spins at arbitrarily large distances will be active, and consequently the resonant percolation cluster will include all of the spin configurations within a given radius $r$ whose energy lies within an exponentially small window of width $\sim \gamma^r$. (This narrow window appears in \cite{Roeck2016} as the hybridization width.)

A related situation was considered in rigorous work by Aizenman and Warzel on resonant delocalization on the Bethe lattice \cite{Aizenman2011,Aizenman2011a,Aizenman2013}.
They show that long-range tunneling to distant resonant sites can lead to delocalization, provided there are sufficiently many paths available, and hence sufficiently many opportunities for sites to be on-resonance to the requisite accuracy.   

\section{Numerical verification} \label{sec: numerical verification}
\begin{figure}[h!]
\begin{center}
\includegraphics[width=7cm,height=2cm]{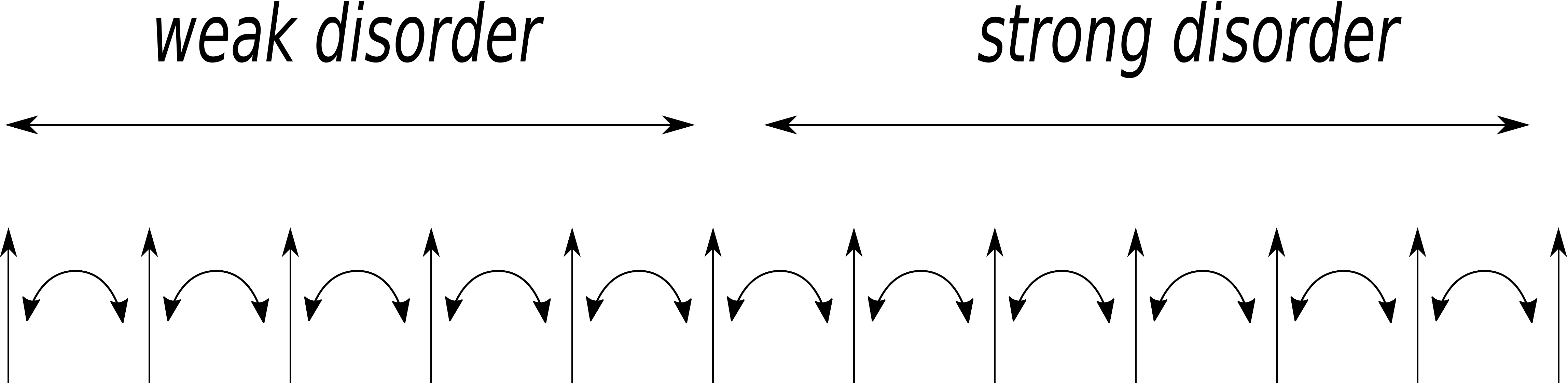}
\\
\includegraphics[width=7cm,height=1.7cm]{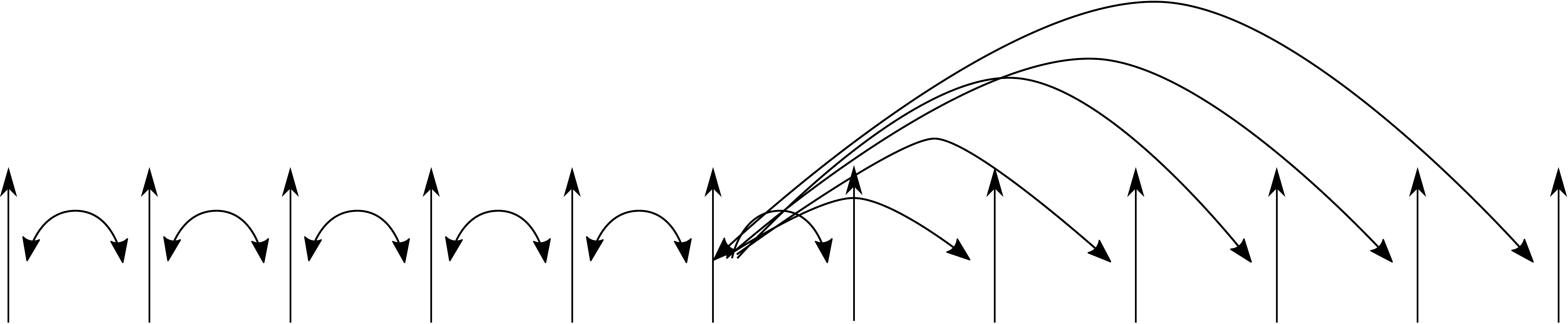}\\
\includegraphics[width=7cm,height=1.7cm]{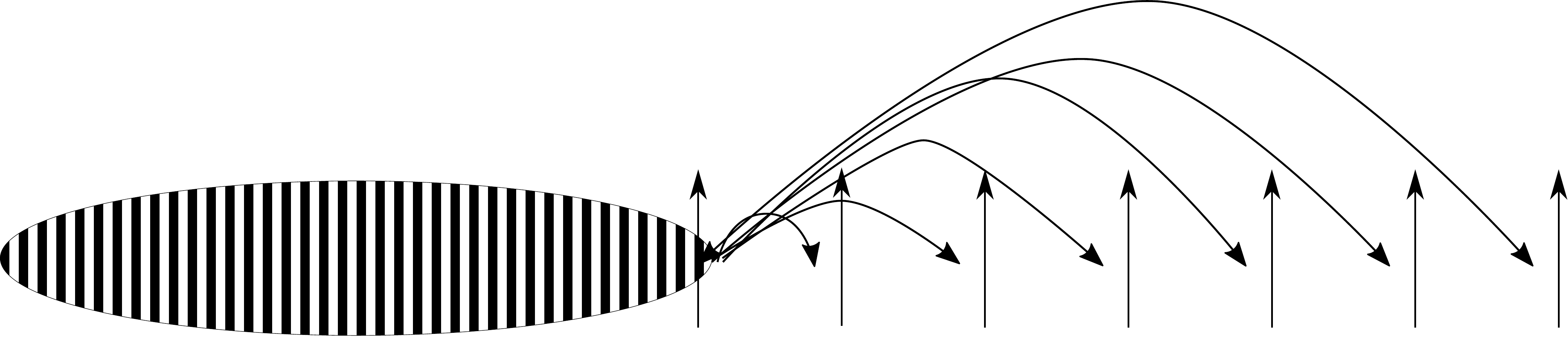}
\end{center}
\caption{
\label{fig: modelling}
\emph{Top}: A chain of weakly disordered spins  -- the Griffiths region -- is coupled to a chain of strongly disordered spins -- the MBL system.
\emph{Middle}: The system is modeled by weakly disordered spins coupled to LIOMs (no coupling between the LIOMs anymore). Note that all LIOMs are coupled only to the rightmost spin.
\emph{Bottom}: The weakly disordered spins are modeled by a random matrix.
}
\end{figure}

The ideas developed above seem amenable to numerical tests, but there are challenges. First of all, we need a Griffiths region to act as a ``seed" of ergodicity. In practice, this means that we need to engineer weak disorder in one region of the spin system. Since it is well-accepted \cite{Srednicki1994,Deutsch1991,Rigol2008,d2016quantum} that ergodic spin systems can, up to some caveats, be modeled by random matrices, it seems well-justified to replace the Hamiltonian in the Griffiths region by a random matrix of the right symmetry class (GOE in our case since we have real Hamiltonians). Hence the model is $H=H_{\text{Gf}}+H_{\text{loc}} +H_{\text{Gf-loc}}$ where $H_{\text{Gf}}$ is the GOE random matrix describing the Griffiths region ``Gf", $H_{\text{loc}}$ describes the localized surroundings ``loc," where we assume that the perturbative procedure described in Section \ref{sec:2} works perfectly, i.e.\ we never encounter any resonance. Finally, $H_{\text{Gf-loc}}$ describes the interaction terms connecting the two regions, i.e.\ located at $\partial \text{Gf}$. As said, the perturbative diagonalization can be done for $H_{\text{loc}}$, which means that there is a quasilocal transformation matrix $U$ such that $UH_{\text{loc}}U^*=  D=D(S^{z}_i, i \in \text{loc}) $. The transformation $O \mapsto U O U^*$ transforms local operators at a site $i$ into operators that decay exponentially with distance from $i$. In particular, after implementing the transformation, the coupling term $H_{\text{Gf-loc}}$ consists of a sum of terms located at sites $i \in \text{loc}$ and with strength decaying exponentially in the distance to the boundary $\partial \text{Gf}$. Finally, we introduce another simplification: The system in the localized region is taken to be non-interacting (Anderson insulator). It is hard to imagine that this would \emph{weaken} the tendency of the total system to localization, and hence this simplification only strengthens our case provided that we still exhibit delocalization. So, finally, we model the resulting Hamiltonian as 
$$
H=  H_{\text{Gf}} +  \sum_{i \in \text{loc}} h_i S^z_{i} +   V_{\text{Gf}} \sum_{i \in \text{loc}} J_i  S^x_{i} ,\qquad  J_i = J_0 \alpha^{\text{dist}(\partial \text{Gf},i)}
$$ 
where $V_{\text{Gf}}$ acts inside the Griffiths region and we see that the couplings $J_i$ decay exponentially when moving away from the Griffiths region.  The absence of terms like $S^z_iS^z_j$ is due to the restriction to an Anderson insulator. 
Let us now fix the geometry of the setup. Inside the Griffiths region, it does not matter since we are anyhow modeling that by a random matrix. The exterior localized region we will take one-dimensional, even though the whole point is to substantiate claims about higher dimensions. The reason for this is that the maximal number of spins that we can reasonably consider is $14-18$ (if we do exact diagonalization, which is the only unbiased method available), making it hard to imagine arranging this small amount of spins credibly in a $d>1$ setup.  The process leading to our model Hamiltonian is illustrated in Fig. \ref{fig: modelling}. 
To analyze our setup, the philosophy is to use criterion \ref{(3.3)} repeatedly. 
When adding the first spin, say $i=1$, the dimension of the bath is $d_{\text{Gf}}$ and the tunneling amplitude due to the term $ J_i V_{\text{Gf}}  S^x_{i}$ is 
$
J_i/\sqrt{d_{\text{Gf}}}
$.
Here we used crucially the random matrix structure of the Griffiths region to estimate a matrix element of the operator $V_{\text{Gf}}$ between eigenstates. 
This tunneling amplitude is to be compared with the level spacing $ W_{\text{Gf}}d_{\text{Gf}}^{-1}$ where $W_{\text{Gf}}$ is the spectral width. Since this grows typically linearly with the size of the Griffiths region, it is completely irrelevant given that $d_{\text{Gf}}$  grows exponentially.  Hence criterion \ref{(3.3)} teaches us that the first spin is thermalized provided that
\begin{equation} \label{eq: criterion concretized}
\frac{J_1 d_{\text{Gf}}}{W_{\text{Gf}}\sqrt{d_{\text{Gf}}}} =  \frac{J_1 }{W_{\text{Gf}}} \sqrt{d_{\text{Gf}}} \gg 1.
\end{equation}

\begin{figure}[h!]
\begin{center}
\includegraphics[width=7cm,height=7cm]{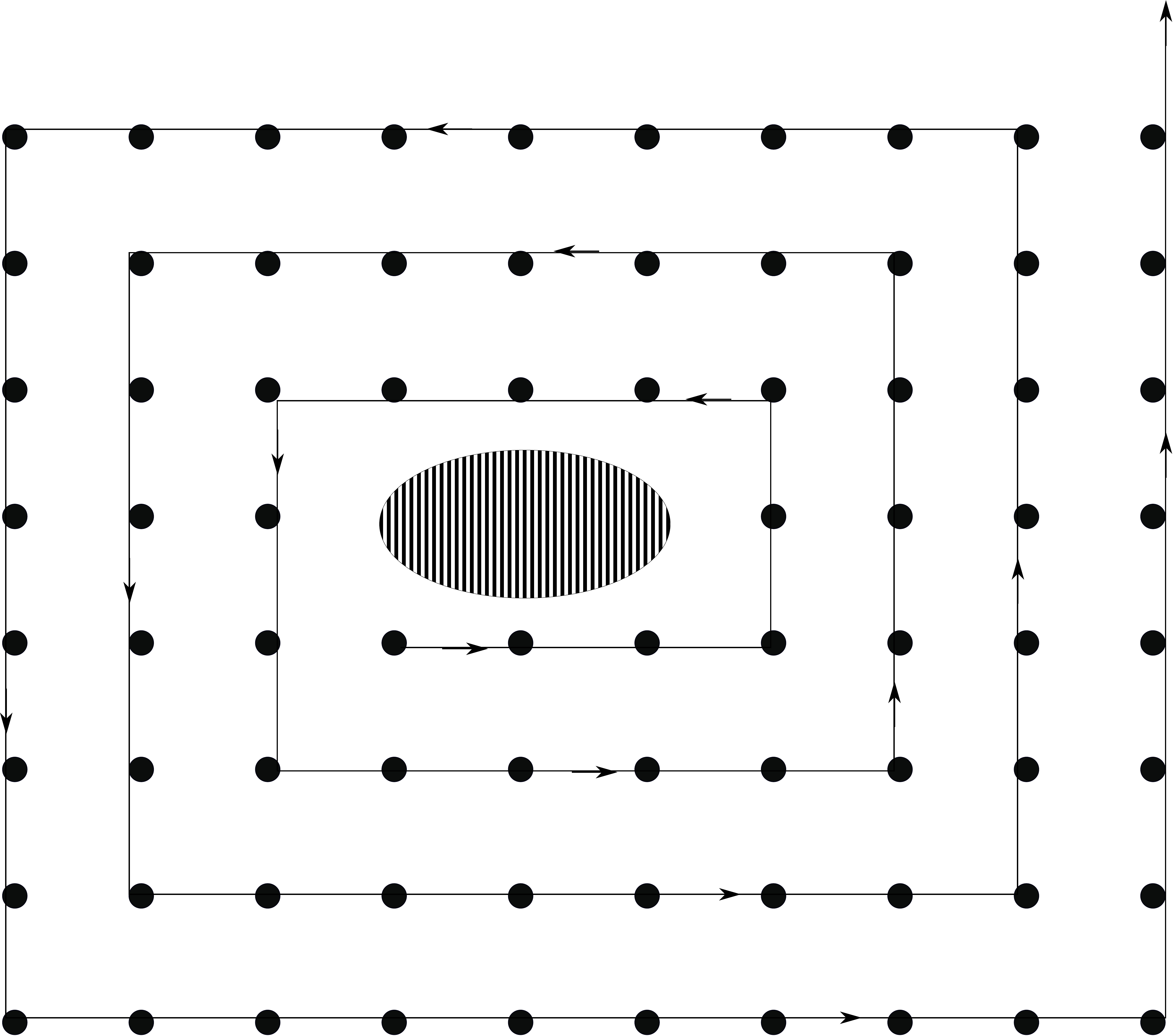}
\caption{
\label{fig: spirals}
A Griffiths region in $d=2$. The distance to the region grows sublinearly with the (somewhat arbitrary) index of the coupled spins, arranged along the spiral.
}
\end{center}
\end{figure}

We see that this is obviously satisfied if the Griffiths region is large enough ($d_{\text{Gf}} \gg 1$).  What happens with the next spins? Since now the bath has been strengthened by the first spin, we have to update $d_{\text{Gf}}\to 2d_{\text{Gf}} $ which makes it easier for the next spin to be thermalized. On the other hand, the coupling is decreased since $J_1\to J_2 =J_1\alpha$. 
Hence, by inspecting \eqref{eq: criterion concretized} we arrive at the following dichotomy:
\begin{enumerate}
\item Either $\alpha <\sqrt{1/2}$, then eventually the bath runs out of steam. More precisely, after coupling  $\ell$ spins with $\ell$ such that
$$
  \frac{J_\ell }{W_{\text{Gf}}} \sqrt{d_{\text{Gf}}2^\ell} \approx 1
   \quad \Leftrightarrow\quad   
          \ell \approx  -\frac{1}{\log(\sqrt{2}\alpha)}  \log(  \frac{J_0 \sqrt{d_{\text{Gf}}} }{W_{\text{Gf}}} )
$$ 
any further spin that is coupled does not get thermalized any more. This value of $\ell$ is hence a prediction for the size of the physical buffer region depicted in Fig. \ref{fig2}. 
\item If $\alpha >\sqrt{1/2}$, then the bath simply gets stronger the more spins are coupled to it. The system is delocalized.
\end{enumerate}
If we would now arrange the spins in a higher dimensional geometry and the index $i$ would spiral away from the Griffiths region, then clearly the couplings $J_i$ decay slower than exponentially in $i$, see Fig. \ref{fig: spirals}.
%
This falls hence automatically in case $2$ above, in line with the discussion of Section \ref{sec:3}. 

The next question is how to diagnose the delocalization of the whole system.  We distinguish three possible tests of the above theory:
\begin{enumerate}
\item Global characteristics. In general, to distinguish between a localized and an ergodic system, we can rely on spectral statistics: we expect level repulsion and random matrix statistics for an ergodic system and absence of level repulsion and Poissonian statistics for a localized system.
\item  Characteristics of the added spins. We can test whether these extremely weakly coupled spins indeed get thermalized by the Griffiths region. This can e.g.\ be done by looking at the distribution of values of local observables located at those spins over distinct eigenvectors. The ETH predicts that all eigenvectors should have similar values.
\item Characteristics of the Griffiths region itself. We can test the effective dimension of the ergodic region by determining what dimension $d_{\text{eff}}$ of random matrix matches best its behavior. The most  natural way to do this seems to be by looking at off-diagonal matrix elements of local operators, leading to a many-body notion of IPRs explained below. 
\end{enumerate}

Numerical tests of the type 1, 2 above are currently under way \cite{luitz2017small} and they seem to confirm the phase transition at $\alpha=\sqrt{1/2}$. Numerical tests of the type 3 have been done in \cite{Roeck2016}, so let us discuss them. 
Let $O$ be an operator in the Griffiths region with $O=O^*$. Then ETH predicts that for eigenstates $\psi,\psi'$ sufficiently close in energy (more precisely, $| E(\psi)-E(\psi') |$ should be no bigger than the Thouless energy), one has
\begin{equation} \label{eq: off diagonal eth}
|\langle \psi, O \psi' \rangle | \sim \frac{1}{d_{\text{eff}}} = e^{-S/2},
\end{equation}
in particular the left-hand side is roughly independent of the precise eigenstates, and $S$ is the entropy at energy $E(\psi) \approx E(\psi')$.
Of course, the off-diagonal elements satisfy a sum-rule since
$$
\sum_{\psi,\psi'}|\langle \psi, O \psi' \rangle |^2= \mathrm{Tr}( OO^*).
$$
 Therefore, averaging the square of matrix elements brings no information and instead we look at a higher moment, defining the $\psi$-dependent \emph{Inverse Participation Ratio} 
 $$
(\mathrm{IPR}_O(\psi))^{-1} :=\sum_{\psi'}  |\langle \psi, O \psi' \rangle |^4. 
 $$
From \eqref{eq: off diagonal eth} we get the interpretation of $\mathrm{IPR}_O(\psi)$ as
$$
\mathrm{IPR}_O(\psi) \approx  d_{\text{eff}}. 
$$
This interpretation as effective dimension ultimately justifies the name ``IPR" as an analogous quantity to the inverse of $\sum_{x} |\psi(x)|^4$ for single-particle wavefunctions.  Just like in the single-particle case, the IPR could be energy-dependent (or even $\psi$-dependent) and we always choose $\psi$ in the middle of the spectrum. 

\begin{figure}[h!]
\begin{center}
\includegraphics[width=9cm,height=5cm]{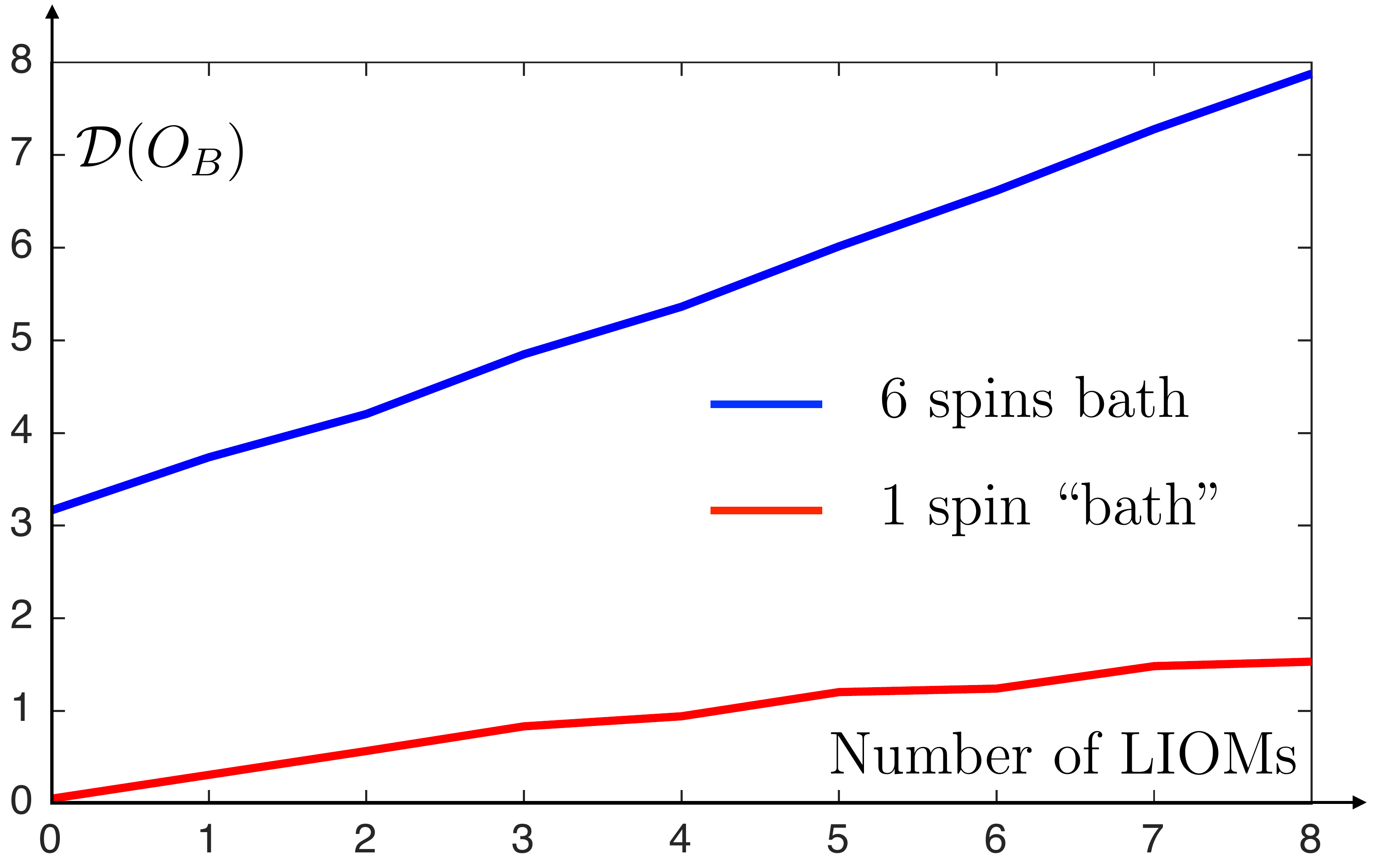}
\end{center}
\caption{
\label{fig: IPR bath}
On the vertical axis we plot $\mathcal D(O_{B})=\langle\ln(\mathrm{IPR}_O(\psi))\rangle$ (average over states near the middle of the spectrum, and over disorder realization). On the horizontal axis, we indicate the number of previously localized spins coupled to the Griffiths region. The blue curve corresponds to a Griffiths region consisting of $6$ spins (modeled by GOE) whereas the red curve corresponds to a powerless Griffiths region of just one spin.  A close look reveals that the blue curve increases nearly $\ln 2$ with each added spin, corresponding to a doubling of the IPR, as described above. If one were to plot the same curve in a setup where $\alpha<\sqrt{1/2}$, one would see the curve level off and become flat after a few added spins, see \cite{luitz2017small}.}
\end{figure}

In the scenario $\alpha> \sqrt{1/2}$ discussed above, we expect that $\mathrm{IPR}_O(\psi)$ increases by a factor of $2$ for each spin that is coupled to the Griffiths region (at least if $O$ itself is in the Griffiths region).  This is indeed confirmed almost perfectly by the numerics, see Fig. \ref{fig: IPR bath}, for the case $\alpha=3/4> \sqrt{1/2}$.  

\section{Critiques}\label{sec:5}

Finally, let us comment on weaknesses of the above arguments.  Regardless of the form in which the ``bootstrap'' argument is phrased, it inevitably relies on the fact that thermalized regions exhibit some chaotic behaviour towards the distant LIOM's that are coupled to it. In the present section \ref{sec: numerical verification}, this was apparent in the fact that matrix elements were computed by a random matrix ansatz.  One can certainly imagine scenarios in which the coupling of the first LIOMs would ``kill'' the bath, so that it would become unable to thermalize more LIOMs. Such scenarios go under the name ``proximity effects'', see \cite{nandkishore2015many,nandkishore2016}, also explored in \cite{hyatt2017many}. The easiest way for this to happen, is if the system localizes in the basis of the bath-LIOM coupling $H_{\text{Gf-loc}}$. However, this requires that $H_{\text{Gf-loc}}$ dominates the term $H_{\text{Gf}}$.  For a realistic model of a Griffiths region, this is not the case, but it can definitely happen in toy models. For example, this will occur if the parameter $\alpha$ introduced above is taken too close to $1$ without reducing $J_0$, since $ || H_{\text{Gf-loc}} || \propto \frac{J_0}{1-\alpha}$ (by summing a geometric series).

Another, perhaps not entirely unrelated concern, is the following:  If  a random matrix assumption is made at each point of the ``bootstrap''
argument, does that not ignore the increasingly long timescales that are present in the system, thus suggesting that the whole procedure is inconsistent?  
The timescale for thermalization may be equated with the inverse of the hybridization width, which as discussed above, decreases exponentially with the distance from the bubble. But the level-spacing decreases more rapidly for $d>1$, so (\ref{(3.6)}) indicates that thermalization extends to all distances, albeit at exponentially small rates. 
However, this argument ignores any potential backreaction of the spins on the bath.
Indeed, if a system is very weakly coupled to an external spin, then the backreaction should be visible in \emph{structure factors} of all local operators as follows: these spectral factors should have  a peak at the precession frequency of the spin, with a very narrow width that depends on the coupling of that spin to the system. After all, the spectral factors are related by Fourier transform to two-time correlation functions (see \cite{d2016quantum} for a detailed discussion of such matters) so that slow degrees of freedom leave their fingerprint on the spectral factors.
Hence,  a theory is surely not completely accurate if it does not allow for any backreaction of the weakly coupled spins on the bath. 
Now,  the theory developed in \cite{Roeck2016} \emph{does} include backreaction effects and, in particular, narrow peaks and troughs in the structure factors of local operators do appear. Yet, the total structure factor still has a dominant continuous background whose weight is larger than that of the fractal part. Of course, the analysis of \cite{Roeck2016} ultimately relies on a model to characterize the new eigenfunctions and hence it cannot claim to settle the issue conclusively. 

Another scenario that is sometimes brought forward, see \cite{banerjee2017solvable}, is that the system would undergo a phase transition when the number of LIOMs becomes comparable to the number of degrees of freedom in the bath.  The microscopic mechanism for such a scenario is not clear to us. 
However, the analysis in \cite{Roeck2016} gives support to the idea that large numbers of LIOMs (or even infinitely many) can be treated perturbatively if the interaction strengths are small in an $\ell^2$ sense.
Indeed, as discussed above, the critical value $\alpha_c = \sqrt{1/2}$ 
is the point at which the effects of decaying interaction strengths balances the effects of increasing numbers of degrees of freedom.

\section{Non-ergodicity or quasi-localization?}\label{sec:6}

Several authors have speculated that for certain delocalized systems, the wavefunctions could yet fail to be non-ergodic, as they are effectively supported on a Hilbert space of much smaller dimension than the full space. See, for example, \cite{Torres2017} and references therein. In particular,  \cite{Altshuler2016} provided numerical evidence for this scenario on random regular graphs. However, other authors  \cite{tikhonov2016anderson} classified this as a finite-size effect, and it seems fair to say that this issue has not been conclusively resolved yet. 
One could try to draw parallels with the many-body problem we have been considering as the network of resonances that eventually delocalizes the system is very sparse. Therefore, the question arises whether our case matches the ``non-ergodic delocalization"-label, at least in the case where there is a single Griffiths region for the entire system. 
Let us stress that it is certainly clear that thermalization in such systems will occur exceedingly slowly. The thermalization time for a system located at distance $r$ from the Griffiths region will grow exponentially with $r$, and hence, when $r$ is of the order of the volume itself, one can consider this system localized for all practical purposes, just as the systems considered in \cite{Chandran2016}. Whether one calls such a system ergodic or not, is largely a matter of convention. It could be ergodic in the sense that it probably follows random matrix statistics (unpublished numerics \cite{luitz2017small} seems to point in that direction), but it is nonergodic in the same sense as finite systems having Ising symmetry breaking, where one has to wait a time of order $\exp (\text{volume})$ before seeing the ``other phase".  
If we restore a finite density of Griffiths regions, as in real materials, than the picture changes: now we expect the thermalization time to be independent of volume, though still ridiculously large. 
This fits well with all the previous work on ``quasi-localization", see e.g.\ \cite{DeRoeck2013,yao2016quasi} where one expects systems to exhibit very slow dynamics, not unlike the dynamics in glasses. The extremely long time scales required for thermalization are presumably not accessible experimentally. Thus there is no contradiction with
experiments on two-dimensional optical glasses indicating a transition to a non-thermalizing MBL phase \cite{kondov2015disorder,choi2016,bordia2017probing}.
\begin{figure}[h!]
\begin{center}

\includegraphics[width=8.5cm,height=5cm]{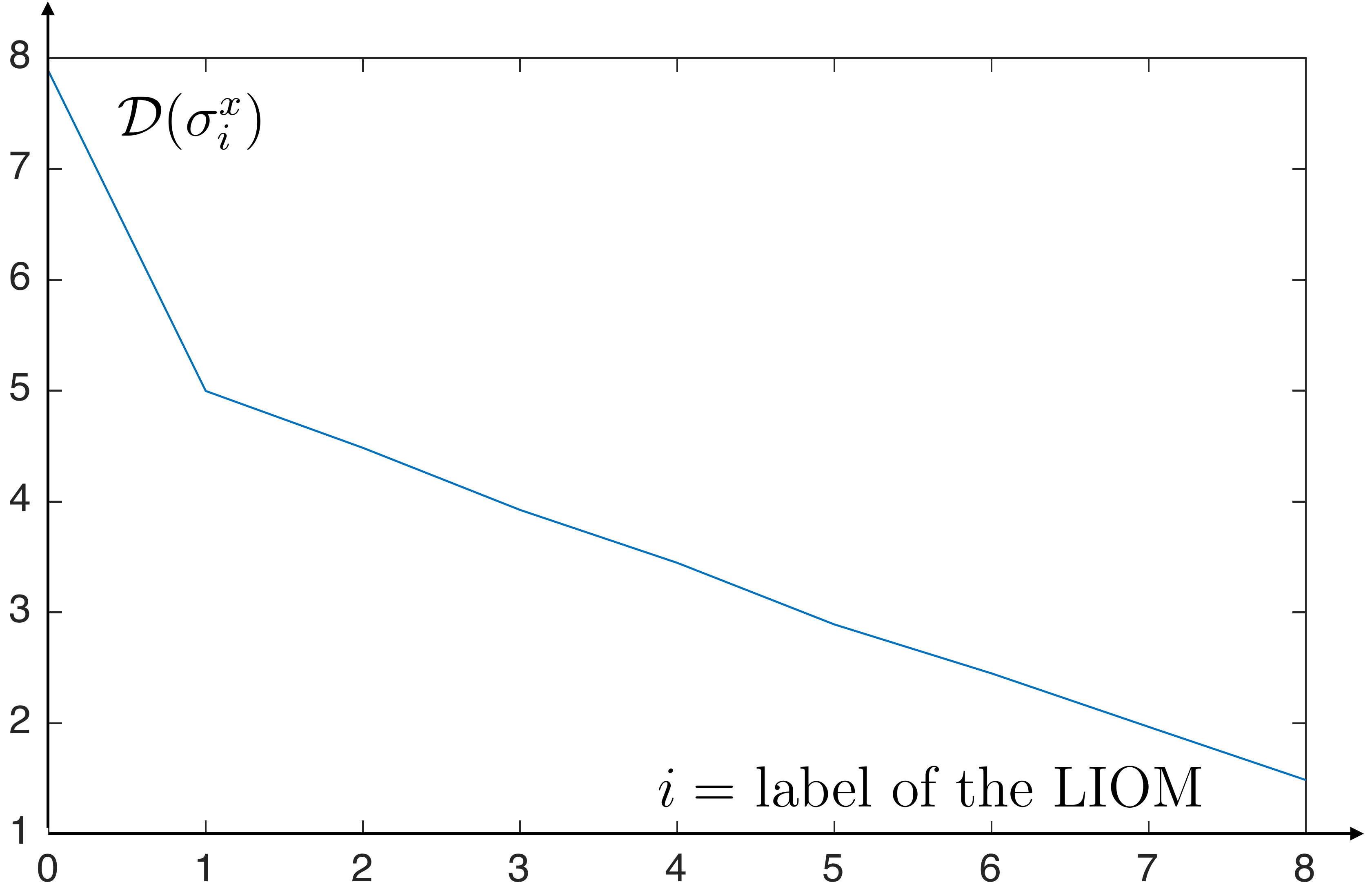}

\end{center}
\caption{
\label{fig: IPR buffer region} 
Same setup as in ref{fig: IPR bath}, but now 
$\mathcal D$ is computed for an operator located at distance $i$ from the Griffiths region. The theory predicts that from $i=1$ onwards,  $\mathcal D$ decreases with slope $-2\log(\alpha)$, which in the case at hand equals $-0.58$. This seems in reasonable agreement with the slope $-0.5$ seen in the plot.}
\end{figure}

However,  what comes very close to a picture of non-ergodicity is the fact that the IPR as defined above, depending on a local operator $O_i$, does decrease when one moves away from the Griffiths region. Roughly, one finds 
$$
\mathrm{IPR}_{O_i}(\psi) \approx  \alpha^{2r} \mathrm{IPR}_{O_{\text{Gf}}}(\psi),
$$
where $r$ is the distance of $O_i$ to the Griffiths region, where $O_{\text{Gf}}$ is located. This exponential dependence has been predicted by the theory of \cite{Roeck2016} and confirmed numerically (see Fig. \ref{fig: IPR buffer region}).  Standard ETH reasoning allows to translate the factor  $ \alpha^{2r}$ into a decrease of a local thermalization time, see e.g.\ \cite{d2016quantum,Roeck2016}.
However, since the same wavefunction $\psi$ is concerned on the left and the right side, the straightforward interpretation as an effective dimension is no longer tenable \emph{literally} and therefore it is not quite clear how to precisely rephrase these findings in terms of the support of typical wavefunctions.

 \section*{Acknowledgement}
 The authors would like to thank A. Chandran and E. Altman for helpful discussions. WDR thanks F. Huveneers and D. Luitz for collaborations on which this review is partially based.   WDR is supported by the FWO (Flemish Research Fund).  
\begin{footnotesize}

\end{footnotesize}
\end{document}